\documentclass[%
 twocolumn,
 aps, pre,
 amsmath,amssymb,
 superscriptaddress
]{revtex4-1}

\usepackage{graphicx}% Include figure files
\usepackage{dcolumn}% Align table columns on decimal point
\usepackage{bm}% bold math
\usepackage[utf8]{inputenc}
\usepackage[T1]{fontenc}
\usepackage{graphicx}% Include figure files
\usepackage{dcolumn}% Align table columns on decimal point
\usepackage{bm}% bold math
\usepackage{xcolor}
\definecolor{blue}{rgb}{0,0.5,1.0}
\usepackage[export]{adjustbox}
\usepackage{braket}
\usepackage{amsmath}
\usepackage{MnSymbol}
\usepackage{bbm}

\begin{document}

%Title of paper
\title{Dissipative nonequilibrium synchronization of topological edge states via self-oscillation}
%Authors

\author{C. W. W\"achtler}
\email{christopher.w.waechtler@campus.tu-berlin.de}
\affiliation{Institute of Theoretical Physics, Technical University Berlin, Hardenbergstra\ss e 36, D-10623 Berlin, Germany.}
\affiliation{NTT Basic Research Laboratories, NTT Corporation, 3-1 Morinosato Wakamiya, Atsugi, Kanagawa 243-0198, Japan.}
\author{V. M. Bastidas}
\affiliation{NTT Basic Research Laboratories, NTT Corporation, 3-1 Morinosato Wakamiya, Atsugi, Kanagawa 243-0198, Japan.}
\author{G. Schaller}
\affiliation{Institute of Theoretical Physics, Technical University Berlin, Hardenbergstr\ss e 36, D-10623 Berlin, Germany.}
\author{W. J. Munro}
\affiliation{NTT Basic Research Laboratories, NTT Corporation, 3-1 Morinosato Wakamiya, Atsugi, Kanagawa 243-0198, Japan.}
\affiliation{National Institute of Informatics, 2-1-2 Hitotsubashi, Chiyoda, Tokyo 101-0003, Japan.}

\date{\today}

%%******Abstract*****%%
\begin{abstract}
The interplay of synchronization and topological band structures with symmetry protected midgap states under the influence of driving and dissipation is largely unexplored. Here we consider a trimer chain of electron shuttles, each consisting of a harmonic oscillator coupled to a quantum dot positioned between two electronic leads. Each shuttle is subject to thermal dissipation and undergoes a bifurcation towards self-oscillation with a stable limit cycle if driven by a bias voltage between the leads. By mechanically coupling the oscillators together, we observe synchronized motion at the ends of the chain, which can be explained using a linear stability analysis. Because of the inversion symmetry of the trimer chain, these synchronized states are topologically protected against local disorder. Furthermore,  with current experimental feasibility,  the synchronized motion can be observed by measuring the dot occupation of each shuttle. Our results open another avenue to enhance the robustness of synchronized motion by exploiting topology.
\end{abstract}

\maketitle

\section{Introduction}\label{sec:Introduction}
Topology plays a fundamental role in several fields ranging from condensed matter physics~\cite{KlitzingEtAlPRL1980, thouless1983,Thouless1984,Seiler1985,KoenigEtAlScience2007,KaneMelePRL2005-2,KaneMelePRL2005}, to gauge field theories in high-energy physics~\cite{1980Daniel,1989witten, Almheiri2015}. In gauge theories, for example, it allows us to understand topological charges such as magnetic monopoles and instantons~\cite{Gross1981}. In condensed matter physics, topology is of utmost importance for our understanding of the integer and fractional quantum Hall effects~\cite{Stormer1999}, topological insulators and superconductors~\cite{HasanKaneRMP2010,QiZhangRMP2011}. Recent experiments in platforms ranging from ultracold optical superlattices~\cite{bloch2016, takahashi2016,2018Lohse} to waveguide arrays~\cite{kraus2012,Silberberg2015,2018Zilberberg} have demonstrated topological protection of transported particles. In mechanical systems, one can exploit topological protection to design materials with desired properties~\cite{Nash2015,Rocklin2017,Mao2018}. While topological systems exhibit robustness against imperfections in closed systems, it is a nontrivial problem to determine if this feature is still available in open systems~\cite{Diehl2011,Bardyn2013,BalabanovJohannesonPRB2017, Gal2020}.  A recent example of this is the theoretical prediction~\cite{Harari2018} and experimental realization~\cite{Bandres2018} of topological lasers \cite{Malzard2018, Smirnova2019}.

Systems coupled to external reservoirs~\cite{BreuerPetruccioneBook2002,CarmichaelBook1993,SchallerBook2014} exhibit a variety of phenomena with no counterpart in closed systems~\cite{Diehl2010,dimitrisreview,Foss2017, Kyaw2020}. One of these is synchronization, which is a hallmark of collective behavior in nonequilibrium systems~\cite{Strogatz2014}.
This phenomenon was first observed by Huygens in 1665 in coupled clocks~\cite{bennett2002huygens} and has been studied in diverse communities since then. Synchronization plays an important role in our understanding of electric networks in engineering, circadian rhythms in biology, pattern formation, and chemical reactions~\cite{Strogatz1993,Rosenblum2003,Arenas2008}. In nonlinear dynamics, synchronization is related to the emergence of collective periodic motion in networks of nonlinear coupled self-sustained oscillators~\cite{Strogatz2014}. In physics, this phenomenon has been extensively explored in classical systems both theoretically as well as experimentally~\cite{Strogatz1993, Rosenblum2003, Arenas2008}. However, its quantum counterpart remains largely unexplored. Recent works have reported synchronization of quantum Stuart-Landau oscillators~\cite{Lee2013,Lee2014,bastidas2015,Walter2015,PhysRevLett.120.163601, mok2020noise, hajduek2020comment, mok2020reply, PhysRevE.101.020201},
arrays of coupled spin systems \cite{Giorgi2013,Hush2015} and other manybody systems \cite{Davis2018,Tindall2020}. In the theory of synchronization, the system usually reaches a steady state that is independent of the initial conditions. In this sense, the system is robust against changes in its initial configuration. However, under perturbations of the system parameters, the synchronized state of the system may change.

\begin{figure}
\begin{center}
\includegraphics[width=\columnwidth]{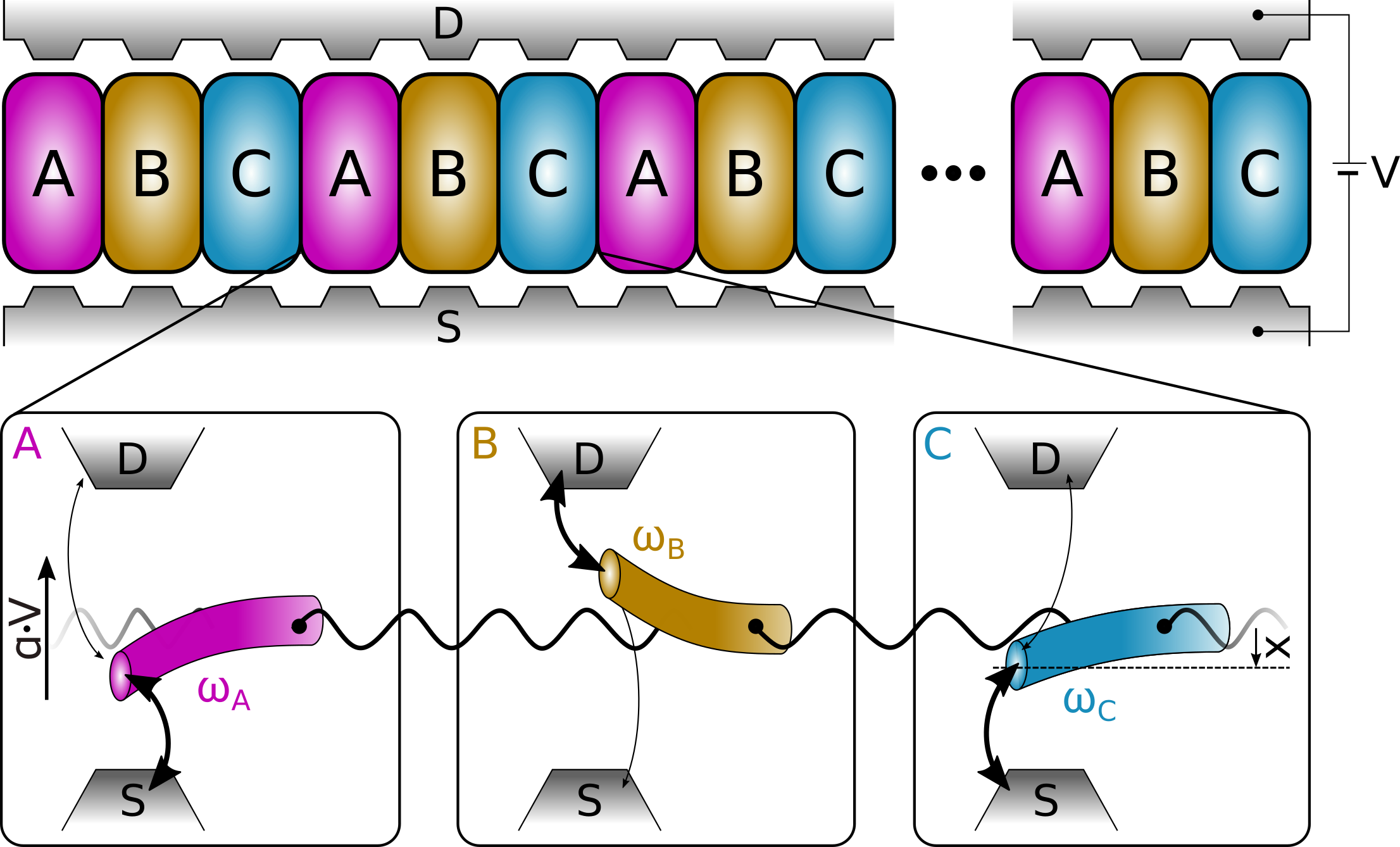}
\end{center}
\caption{Dissipative nonequilibrium system with topological edge states: An open chain of $N$ coupled systems splits into trimers with units $A$, $B$ and $C$. Each system consists of an electron shuttle, that is a quantum dot (QD) hosted by a nanomechanical oscillator positioned between electronic leads. The QD is tunnel coupled to the leads, source (S) and drain (D), in a nonlinear fashion such that electron tunneling is enhanced in the proximity of the lead (indicated by the thickness of the arrows). The same bias voltage $V$ is applied to each shuttle and induces an electrostatic field $\alpha V$ between the leads, which in turn forces the charged shuttle to move toward the drain. Each shuttle is coupled to its nearest neighbors via the mechanical coupling of the oscillators to build a chain. Trimerization of the chain is then achieved by assigning different frequencies to nanomechanical oscillators of each unit, e.g. by varying oscillator length}
\label{fig:Model}
\end{figure}

In our work, we exploit topological protection to enhance the robustness of synchronization phenomena in quantum transport. A key ingredient to achieve synchronized motion is self-oscillation. The latter naturally appears in the electron shuttle, a paradigmatic model of quantum transport \cite{GorelikEtAlPRL1998, IsacssonEtAlPhysicaB1998, WeissZwergerEPL1999, BoeseSchoellerEPL2001, NordEtAlPRB2002, ArmourMacKinnonPRB2002, FedoretsEtAlEPL2002, MccarthyEtAlPRB2003, NovotnyEtAlPRL2003, NovotnyEtAlPRL2004, DonariniEtAlNJP2005,UtamiEtAlPRB2006, UtamiEtAlPRB2006, NoceraEtAlPRB2011, WaechtlerEtAlNJP2019, WaechtlerPRAppl2019}. There, the interplay between sequential electron tunneling  and mechanical motion of the oscillator leads to self-sustained oscillations. Thus, an array of coupled electron shuttles (see Fig.~\ref{fig:Model}) serves as a natural platform to investigate synchronized motion. In contrast to previous studies \cite{BoehlingEtAlPRB2018}, we stress that in our model the electron transport is transversal to the chain. Inspired by models of condensed matter physics such as Aubry-Andr\'e \cite{Aubry1980,kraus2012} and Harper-Hofstadter \cite{Hofstadter1976} models, we modulate the frequencies of the oscillators in space in order to define topological band structures in our system. In this way, topology allows us to synchronize the edge states so that they oscillate with the same frequency, even under the effect of imperfections as long as they preserve the symmetry of the model. Furthermore, we show that signatures of edge state synchronization are observable by measuring the local dot occupation at the edges. 

\emph{Outline:} Our paper is arranged as follows: We start by introducing our model. To this end, in Sec.~\ref{subsec:Shuttle} we first review the single electron shuttle. In Sec.~\ref{subsec:Topology} we discuss the topology of a trimer chain of classical harmonic oscillators before combining the aforementioned concepts into a trimer chain of electron shuttles in Sec.~\ref{subsec:Chain}. We then discuss different synchronization scenarios present in the system and discuss the local dot occupation as a witness of the synchronization in Sec.~\ref{sec:Synchronization}.  Subsequently, in Sec.~\ref{sec:LinearStability}, we explain the synchronization using a linear stability analysis. Furthermore, in Sec.~\ref{sec:Disorder} we demonstrate that the synchronization is topologically protected against disorder if the symmetry of the trimer chain is preserved. Lastly, we summarize and conclude our findings in Sec.~\ref{sec:Conclusions}.

\section{Model}\label{sec:Model}
In our work, we investigate the interplay between topology and synchronization. 
The first key ingredient to define synchronization is the phenomenon of self-oscillation. Motivated by this, in Sec.~\ref{subsec:Shuttle} we consider the electron shuttle. 
The latter exhibits self-oscillation due to the interplay of sequential electron tunneling and mechanical motion of the oscillator. 
Before discussing topology in the open system, we review in detail topological phases appearing in a trimer chain of (classical) oscillators in Sec.~\ref{subsec:Topology}. 
With these elements at hand, in Sec.~\ref{subsec:Chain} we couple several electron shuttles to achieve synchronized motion and define topological protection by modulating the onsite energies of the oscillators.

\subsection{Single electron shuttle}
\label{subsec:Shuttle}
The shuttle device considered here consists of a quantum dot (QD) with on-site energy $\varepsilon$ hosted by nanomechanical oscillator with frequency $\omega$. The QD is tunnel coupled to two isothermal leads, source (S) and drain (D), with chemical potentials $\mu^\text S = \varepsilon + V/2$ and $\mu^\text D = \varepsilon - V/2$, respectively. 
Here, $V=\mu^\text S-\mu^\text D$ denotes the applied bias voltage between S and D. 
Throughout this work, we consider units such that the electron charge is $e \equiv 1$.
Further, we focus on the strong Coulomb blockade regime, where the number of charges in the QD can only take the values $0$ and $1$~\cite{GiaeverZellerPRL1968, KulikShekhter1975, AverinLikharevJLTP1986} and we denote the probability of having one electron by $q$. Thus, $q$ is equivalent to the average charge in the shuttle and its average state is fully described by the triple $\mathbf x = (x,p,q)$, where $x$ and $p$ denote the (mass-weighted) position and momentum of the oscillator, respectively. 
The applied bias voltage $V$ induces an electrostatic force $F^\text{el}_l=\alpha V q$ on the charged shuttle. 
Here, $\alpha$ is an effective inverse distance between the two leads.
Additionally, we consider damping of the nanomechanical oscillator with friction coefficient $\gamma$. 
Hence, the total system is dissipative, through friction and electron tunneling, and driven out of equilibrium by the chemical gradient between source and drain.
For simplicity, we assume that the source and the drain have the same temperatures.
A depiction of the single electron shuttle is shown in the boxes $A$, $B$, and $C$ of Fig.~\ref{fig:Model}.

In this work, we focus on the classical limit of the electron shuttle, sometimes referred to as classical shuttling of particles \cite{ShekhterEtAlNano2013}, which is justified for specific experimental realizations \cite{MoskalenkoEtAlPRB2009}. 
In this limit, we can model the dynamics by a system of coupled nonlinear ordinary differential equations \cite{WaechtlerEtAlNJP2019},
\begin{equation}
\label{eq:NonLinearSingleShuttle}
\left(\begin{array}{c} 
\dot x \\
\dot p \\
\dot q 
\end{array}\right)=
\left(\begin{array}{c}
	p\\
	-\omega^2 x - \gamma p +\alpha V q\\
	 -\Gamma_\text{out}\left(x\right)q +  \Gamma_\text{in}\left(x\right)\left(1-q\right)
\end{array}\right),
\end{equation}
where $\Gamma_\text{out}(x)$ and $\Gamma_\text{in}(x)$ are the tunneling rates from and to the QD, respectively. These are exponentially sensitive to the position $x$ of the oscillator \cite{GorelikEtAlPRL1998, LaiEtAlJoP2012, LaiEtAlJoP2013, FedoretsEtAlPRL2004, FedoretsEtAlEPL2002}, motivated by the exponential sensitivity of quantum mechanical tunneling. The explicit form of the tunneling rates is as follows
\begin{equation}
\label{eq:Rates}
\begin{aligned}
\Gamma_\text{out}\left(x\right) &= \Gamma e^{-x/\lambda}\left[1-f^{\text S}\left(\bar \varepsilon\right)\right] +\Gamma e^{x/\lambda}\left[1-f^{\text D}\left(\bar \varepsilon\right)\right],\\
\Gamma_\text{in}\left(x\right) &=\Gamma e^{-x/\lambda}f^{\text S}\left(\bar \varepsilon\right)+\Gamma e^{x/\lambda}f^{\text D}\left(\bar \varepsilon\right),\\
\end{aligned}
\end{equation}
where $\lambda$ is a (mass-weighted) characteristic tunneling length, $\bar\varepsilon = \varepsilon-\alpha V x$ denotes the charging energy of the filled QD, and $\Gamma$ denotes a bare tunneling rate. 
The Fermi function $f^\nu(\bar\varepsilon) = \left[\exp\left(\beta(\bar\varepsilon-\mu^\nu)\right)+1\right]^{-1}$ of lead $\nu \in\{\text S,\text D\}$ with inverse temperature $\beta$ and chemical potential $\mu^\nu$ denotes the probability of having an electron with energy $\bar\varepsilon$ in lead $\nu\in\{\text S,\text D\}$.
The dynamics described by Eq.~(\ref{eq:NonLinearSingleShuttle}) resembles a mean field description \cite{WaechtlerEtAlNJP2019}, where the mechanical motion is described purely deterministically. 
However, the state $q$ of the QD is represented by a probability and behaves stochastically. Moreover, the underlying mechanism of electron tunneling via Fermi's golden rule is intrinsically quantum. 

The tunneling rates $\Gamma_\text{in}(x)$ and $\Gamma_\text{out}(x)$ in Eqs.~(\ref{eq:NonLinearSingleShuttle}) and (\ref{eq:Rates}) depend \emph{nonlinearly} on the shuttle position $x$.
If the shuttle is close to the source ($x<0$), the probability of electron tunneling between the source and the QD is exponentially enhanced while it is exponentially suppressed between the QD and the drain.
Conversely, if $x>0$ the tunneling between QD and drain is enhanced.
Above a critical value of the applied bias voltage, the interplay of electronic tunneling and mechanical motion of the oscillator leads to a bifurcation of the system into a stable stable limit cycle with periodic motion of the oscillator \cite{GorelikEtAlPRL1998, IsacssonEtAlPhysicaB1998, WeissZwergerEPL1999, BoeseSchoellerEPL2001, NordEtAlPRB2002, ArmourMacKinnonPRB2002, FedoretsEtAlEPL2002, MccarthyEtAlPRB2003, NovotnyEtAlPRL2003, NovotnyEtAlPRL2004, DonariniEtAlNJP2005,UtamiEtAlPRB2006, UtamiEtAlPRB2006, NoceraEtAlPRB2011, WaechtlerEtAlNJP2019, WaechtlerPRAppl2019}.
Since the energy supply from the chemical gradient lacks a corresponding periodicity, the electron shuttle is a good example of self-oscillation with stable periodic motion  without active regulation from the outside \cite{JenkinsPR2013}. We would like to remark that self-oscillation is inherently a phenomenon of nonlinear dynamics, i.e., an equivalent model with linear tunneling rates would not exhibit self-sustained oscillations.

\subsection{Topology of a trimer chain}\label{subsec:Topology}
Symmetry-protected phases constitute one of the most striking applications of topology in physics~\cite{KlitzingEtAlPRL1980, thouless1983,Thouless1984,Seiler1985,KoenigEtAlScience2007,KaneMelePRL2005-2,KaneMelePRL2005, HasanKaneRMP2010,QiZhangRMP2011}. 
One of the first examples of this is the integer quantum Hall effect, where the conductivity of a two-dimensional electron gas is  proportional to a topological invariant~\cite{KlitzingEtAlPRL1980, Seiler1985}. 
By varying the magnetic field, there are  transitions between a series of topological phases characterized by different values of a topological invariant known as the Chern number~\cite{KlitzingEtAlPRL1980, Laughlin1981, Seiler1985}. 
In this case, the topological protection emerges as a consequence of broken time-reversal symmetry, due to the external magnetic field ~\cite{HasanKaneRMP2010,QiZhangRMP2011}. 
In one spatial dimension, the Aubry-Andr\'e model describes a system of particles with modulated on-site energies in space $\omega_l = \Delta\left[2+\cos\left(2\pi l b + \phi\right)\right]$, where $l$ denotes the site index and $\Delta$ is the frequency amplitude~\cite{Aubry1980,kraus2012}. 
Furthermore, $b$ is a real number related to the magnetic field in the integer quantum Hall effect and $\phi$ is a parameter that acts as a synthetic dimension~\cite{kraus2012, Seiler1985}. 
When $b$ is rational, the system exhibits a finite number of bands. 
For example, in the case of $b=1/3$, the system has three energy bands with different topological numbers and in the case of $b=1/2$, the system is connected to the Su-Schrieffer-Heeger model \cite{SuEtAlPRL1979, HeegerEtAlRMP1988}.
The Aubry-Andr\'e model is closely related to the two-dimensional integer quantum Hall effect and inherits topological properties associated to two-dimensional topological invariants \cite{kraus2012}.

\begin{figure}
\begin{center}
\includegraphics[width=\columnwidth]{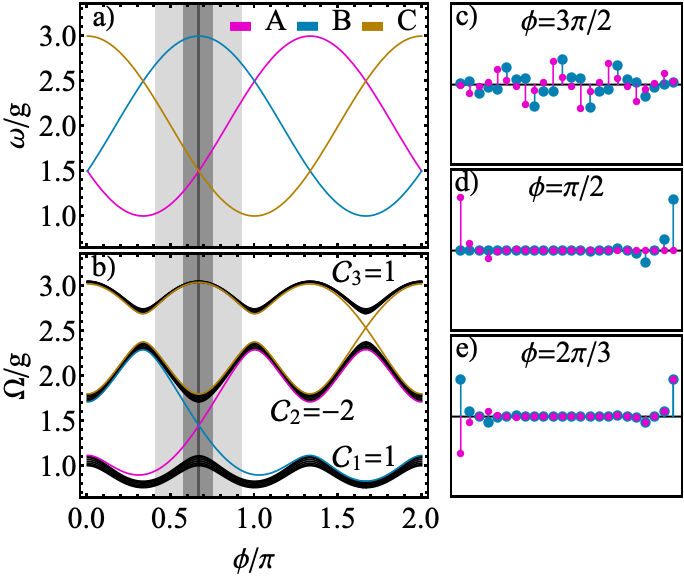}
\end{center}
\caption{(a) Frequencies $\omega_l$ of units $A$ (pink), $B$ (yellow), and $C$ (blue) as function of the global phase $\phi$. (b) Frequency spectrum of $\bm \omega$ [see Eq.~(\ref{eq:HamiltonianTopology})] for $N=24$ and $\Delta/g = 1$. At the crossings in the spectrum ($\phi=2\pi/3$ and $\phi = 5\pi/3$) the Hamiltonian is inversion-symmetric with topologically protected edge states. [(c)--(e)] Corresponding eigenstates of the pink and blue branch in panel (b) for different values of $\phi$ indicating bulk states ($\phi = 3\pi/2$), edge states located at one side of the chain ($\phi = \pi/2$), or inversion symmetric edge states located at both sides of the chain ($\phi=2\pi/3$). The behavior of the orange states is similar (not shown).}
\label{fig:Fig2}
\end{figure}

Motivated by the Aubry-Andr\'e model \cite{Aubry1980,kraus2012}, our aim here is to define topological properties of classical mechanical oscillators \cite{Ma2019}. 
We consider a one-dimensional chain of $N$ units, each consisting of a harmonic oscillator with frequency $\omega_l$.  
The chain is trimerized into units $A$, $B$, and $C$ by choosing $b=1/3$. 
Adjacent units can exchange excitations via mechanical coupling with strength $g$ between neighboring oscillators.
In Fig.~\ref{fig:Fig2}(a), we plot the frequencies $\omega_l$ of the units $A$ (pink), $B$ (yellow), and $C$ (blue) as function of the global phase $\phi$. 
The system of coupled oscillators is described by the Hamiltonian
\begin{equation}
\label{eq:HamiltonianTopology}
H_{\text{osc}} (\phi)= \frac{1}{2}\mathbf p^\intercal \mathbf p + \frac{1}{2}\mathbf x^\intercal  \bm \omega  \mathbf x = \frac{1}{2}\bar{\mathbf p}^\intercal\bar{\mathbf p} + \frac{1}{2}\bar{\mathbf x}^\intercal  \mathbf \Omega  \bar{\mathbf x}, 
\end{equation}
where $\mathbf x \equiv (x_1, x_2, ..., x_N)^\intercal$ and $\mathbf p\equiv (p_1, p_2, ..., p_N)^\intercal $ are mass-weighted position and momentum vectors, respectively. 
Furthermore, $\bm \omega$ is a tridiagonal matrix, where the main diagonal is given by $\text{diag}(\omega_1^2, \omega_2^2,...,\omega_N^2)$ and the upper and lower diagonal by $\text{diag}(-g^2, ...,-g^2)$. 
Hamiltonian (\ref{eq:HamiltonianTopology}) can be diagonalized by a normal mode transformation $O$, such that $\mathbf \Omega=O\bm\omega O^\intercal= \text{diag}(\Omega_1^2, \Omega_2^2,...,\Omega_N^2)$. 
The collective modes are given by $\bar{\mathbf x} = O\mathbf x$ with collective momentum $\bar{\mathbf p} = O\mathbf p$. 

We plot the frequency spectrum $\Omega_l$ of $\bm \omega$ for $N=24$ sites as function of the global phase $\phi$ in Fig.~\ref{fig:Fig2}(b).
The spectrum consists of three dispersive bands with two band gaps. 
Two exemplary bulk states, which are delocalized across the chain, are shown in Fig.~\ref{fig:Fig2}(c), corresponding to the middle (pink) and lower (blue) bands of the spectrum shown in Fig.~\ref{fig:Fig2}(b) for $\phi = 3\pi/2$.
However, as the global phase $\phi$ is changed, these band gaps can host edge states located at either end of the chain [see Fig.~\ref{fig:Fig2}(d)].
For the special cases of $\phi=2\pi/3$ and $\phi=5\pi/3$ with  exponentially small avoided midgap crossings in the frequency spectrum, we find a symmetric and an anti-symmetric edge state localized at both ends of the chain as shown in Fig.~\ref{fig:Fig2}(e) for the case of $\phi=2\pi/3$. 

The above model describes an effective two-dimensional system, where $\phi$ is seen as an additional dimension to spatial dimension \cite{RosaEtAlPRL2019, AlvarezCoutinhoPRA2019}.
Therefore, the topology can be studied using concepts known from two-dimensional (2D) systems. 
Assuming periodic boundary conditions on the system and performing a Fourier transformation of the positions, we can write the Hamiltonian as
\begin{equation}
H_\text{osc}(\phi) = \frac{1}{2} \mathbf p^\intercal \mathbf p + \sum\limits_k \mathbf x^\intercal(k,\phi) \tilde{\bm \omega}(k,\phi) \mathbf  x(-k,\phi) ,
\end{equation}
with Bloch modes $\mathbf x(k,\phi)$. 
For $b=1/3$, there are three bands and three Bloch modes $\mathbf x_n(k,\phi)$.
The Chern numbers of the individual ($n$th) bands can then be defined over the Brillouin zone ($0\leq k<2\pi b$, $0\leq \phi <2\pi$) with $b=1/3$ as
\begin{equation}
\mathcal C_n = \frac{1}{2\pi \imath}\int\limits_0^{2\pi/3} dk\int\limits_0^{2\pi}d\phi \left(\partial_k A^n_\phi -\partial_\phi A^n_k\right),
\end{equation}
where $A^n_\mu = \mathbf x_n(k,\phi)\partial_\mu \mathbf x_n(k,\phi)$. 
We compute the Chern numbers numerically \cite{FukuiEtAlJPSJ2005} and find $\mathcal C_1 = \mathcal C_3 = 1$ and $\mathcal C_2 = -2$ (see Fig.~\ref{fig:Fig2}), such that $\sum_n \mathcal C_n = 0$. 
Furthermore, $\tilde{\bm \omega}(k,\phi)$ is inversion symmetric for $\phi=2\pi/3$ and $\phi=5\pi/3$, that is, $\mathcal P \tilde{\bm \omega}(k,2\pi/3)\mathcal P^{-1}=\tilde{\bm\omega}(-k,2\pi/3)$, where
\begin{equation}
\mathcal P = \left(\begin{array}{ccc}
0 & 0 & 1 \\
0 & 1 & 0 \\
1 & 0 & 0 
\end{array}\right).
\end{equation}
This corresponds to a complete inversion of the chain.
Hence, only the edge states located at \emph{both} ends of the chain are topological as the inversion symmetry is preserved. 
These states are protected against local imperfections as long as the symmetry is preserved.

\subsection{Trimer chain of electron shuttles}\label{subsec:Chain}
So far, we have discussed the onset of self-oscillation in the electron shuttle and how topology can be defined for systems of coupled oscillators. 
Now we have all the ingredients to define topology in synchronized systems. 
We first begin by coupling several electron shuttles, which allows us to synchronize them. 
Afterward, motivated by the  Aubry-Andr\'e model and the discussion in the previous section, we modulate the energies of the oscillators to define topological protection. 
In fact, due to the nature of the electron shuttle, the system is inherently out of equilibrium and the challenge here is to investigate topological protection in non-Hamiltonian systems.

The full model we are interested in consists of a chain of $N$ electron shuttles labeled by the index $l$.
The state of the chain is fully described by the vector $\mathbf X$ with entries $\mathbf X_l = (x_l,p_l,q_l)$ describing the state of the electron shuttle $l$. 
One can model the dynamics of the chain by a differential equation $\dot{\mathbf X} = \mathbf f(\mathbf X)$, where every unit $l$ is governed by the nonlinear equations [see Eq.~(\ref{eq:NonLinearSingleShuttle})]
\begin{equation}
\label{eq:Dynamics}
\dot{\mathbf X}_l = f_l(\mathbf X) = \left(\begin{array}{c}
p_l\\
-\omega_l^2 x_l - \gamma p_l +\alpha V q_l + g^2 \left(x_{l-1} + x_{l+1}\right)\\
 -\Gamma_\text{out}\left(x_l\right)q_l +  \Gamma_\text{in}\left(x_l\right)\left(1-q_l\right)\end{array}\right).
\end{equation}
Here, the electron tunneling rates $\Gamma_\text{in}(x_l)$ and $\Gamma_\text{out}(x_l)$ are defined as in Eq.~(\ref{eq:Rates}).
To fulfill the open boundary conditions, it must hold that $x_{0} = x_{N+1}\equiv 0$. 
Note that the different electron shuttles only differ by the assigned oscillator frequency $\omega_l$.
Hence, Eq.~(\ref{eq:Dynamics}) is inversion symmetric for $\phi = 2\pi/3$ and $\phi = 5\pi/3$ inherited from the inversion symmetry of $H_\text{osc}(\phi)$.
We apply the same bias voltage $V$ to all shuttles.

From the previous section, we know that the spatial modulation of the oscillators leads to topology. 
As a matter of fact, the energies of the collective modes form bands and determine the topology of the system. 
In the system of coupled electron shuttles, the friction as well as the bias voltage is local in space. 
Therefore, to study nonequilibrium effects on the topology, it is convenient to perform a transformation of the equations of motion in terms of collective modes.
Hence, we apply the same normal mode transformation $O$ with entries $O_{lk}$ that diagonalizes $H_\text{osc}(\phi)$ defined in Eq.~(\ref{eq:HamiltonianTopology}). 
In terms of the collective coordinates $\bar{\mathbf X} = (\bar{\mathbf x},\bar{\mathbf p},\bar{\mathbf q})$, where $\bar{\mathbf x}=O\mathbf x$ and similarly for the collective momentum and charge, Eq.~(\ref{eq:Dynamics}) takes the form
\begin{equation}
\label{eq:DynamicsNormalMode}
\dot{\bar{\mathbf X}}_l = %\bar f_l(\bar{\mathbf X}) = 
\left(\begin{array}{c}
\bar p_l\\
-\Omega_l^2 \bar x_l - \gamma \bar p_l +\alpha V \bar q_l\\
 -\Gamma_\text{out}\left[\left(O^\intercal\bar{\mathbf x}\right)_l\right]\bar q_l +  \Gamma_\text{in}\left[\left(O^\intercal\bar{\mathbf x}\right)_l\right]\left(O_l -\bar q_l\right)\end{array}\right).
\end{equation}
Here, $O_l = \sum_k O_{lk}$. Hence, under the coordinate change, the collective modes are now only dissipatively coupled, because the tunneling rates depend on all the collective positions $\bar{\mathbf x}$.

\section{Dissipative nonequilibrium synchronization of topological edge states}\label{sec:Synchronization}
The model introduced in the previous section describes a dissipative nonequilibrium system: 
By applying a bias voltage $V$ between the source and drain of each shuttle, the system is driven out of equilibrium. 
However, the tunneling of electrons as well as the friction of each oscillator accounts for dissipation, such that a stable nonequilibrium steady state can emerge. 
Since each single-electron shuttle undergoes a transition into a stable limit cycle, it is expected that also collective modes of the coupled chain of these shuttles can be excited. 
In fact, we find that for a sufficiently large $V$ and depending on the global phase $\phi$ synchronized states may appear, which persist in the long time limit.
In the following, we only discuss (periodic) steady states, that is, after initial relaxation.

\begin{figure}
\begin{center}
\includegraphics[width=\columnwidth]{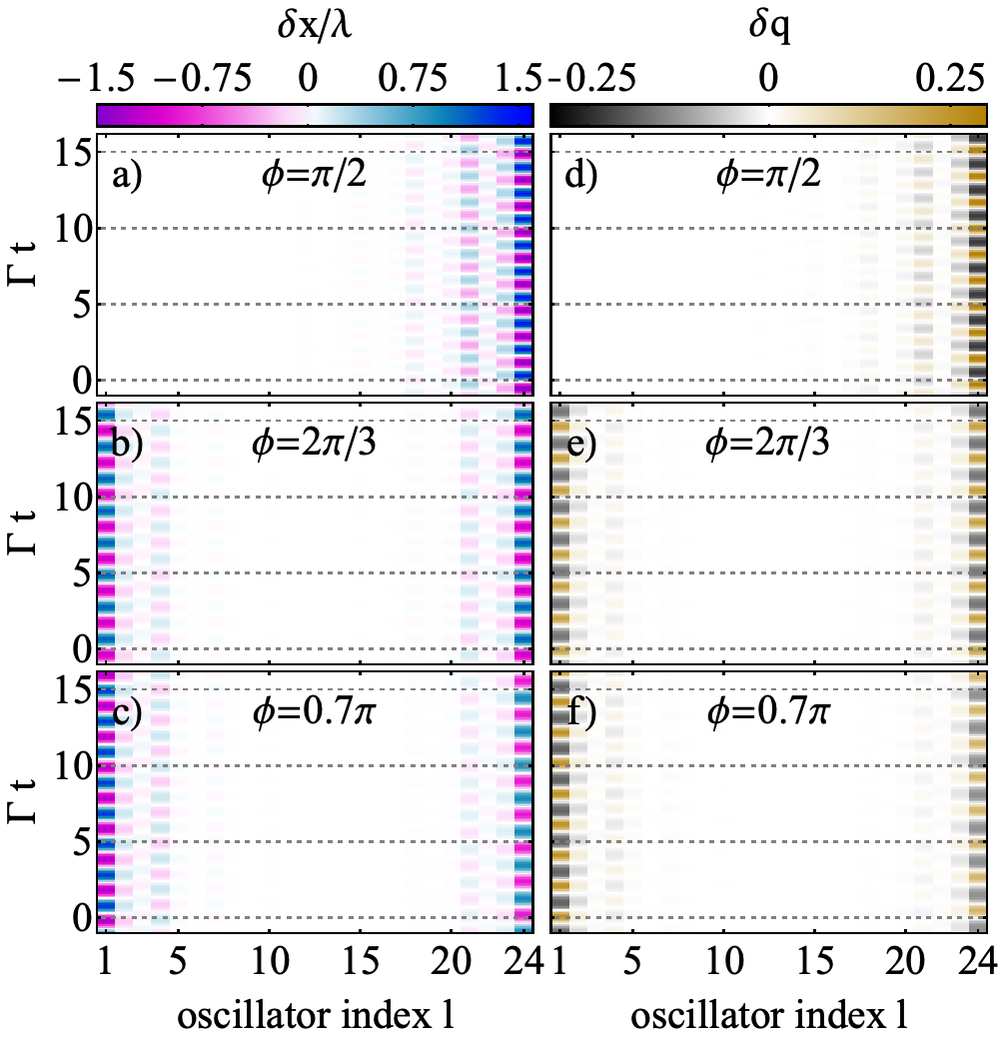}
\end{center}
\caption{Synchronization of a trimer chain of electron shuttles for $N=24$. [(a)--(c)] Oscillating position $\delta x_l(t)$ of the  shuttles, [(d)--(f)] corresponding dot occupation $\delta q_l(t)$ at steady state after initial relaxation. For $\phi = \pi/2$ (top figures), only the right end of the chain performs synchronized motion while the left end is at rest. Consequently, only the dot occupation on the right end of the chain oscillates in time. For $\phi = 2\pi/3$ (middle figures) the inversion symmetry is preserved and both edges of the chain are synchronized with an inversion symmetric initial state. For $\phi = 0.7\pi$ also, both edges oscillate; however, each side of the chain oscillates with its own frequency, such that only shuttles on either side of the chain are synchronized. Parameters: $\beta V = 150.0$, $\alpha \lambda = 0.06$, $\Delta / g = 1.0$, $\Gamma / \gamma = 1.0$.}
\label{fig:Fig3}
\end{figure}

Besides the trivial scenario ($\phi < 0.41\pi$ and $\phi > 0.92\pi$), in which all collective modes are damped, there are three different dynamical scenarios present in the system, for which we show examples in Figs.~\ref{fig:Fig3}(a)--\ref{fig:Fig3}(c). 
To visualize the synchronization of shuttles, we show the deviation $\delta x_l(t)$ of the shuttle position from its time-averaged position, that is, $\delta x_l(t) = x_l(t) - \langle x_l\rangle$.
Here, the averaged position $\langle x_l\rangle = \int_0^{T_l} x_l(t) dt / T_l$ with period $T_l$ of the $l$th shuttle.
Hence, $\delta x_l(t)=0$ implies that the shuttle is not moving over time; however, $\langle x_l\rangle$ may be unequal to $0$.

If $\phi\in [0.41\pi,0.58\pi)$, only the shuttles on the right end of the chain chain oscillate with the same frequency and a fixed phase difference. 
Thus, the right end of the chain performs synchronized motion while the left end of the chain is at rest. 
Moreover, the amplitude decreases exponentially toward the bulk, which is a direct consequence of the existence of the edge states in the system of trimerized oscillators.
We show this situation for the case of $\phi = \pi/2$ in Fig.~\ref{fig:Fig3}(a) as an example.
The opposite behavior, where only the left end of the chain performs synchronized motion, can be observed for $\phi\in (0.75\pi,0.92\pi]$. 

If the trimer chain is inversion symmetric ($\phi=2\pi/3$), both ends of the chain are fully synchronized and the oscillation amplitude again decreases exponentially toward the bulk [see Fig.~\ref{fig:Fig3}(b)].
While the phase difference between the units belonging to the same edge is independent of the initial conditions, the phase between the two ends of the chain does show such a dependency:
A fully inversion symmetric or inversion antisymmetric initial state of the shuttles will be preserved such that the phase difference is $0$ and $\pi$, respectively. 
On the other hand, random initial conditions will lead to a superposition of the inversion symmetric and inversion antisymmetric state with random phase difference. 
However, after initial relaxation the units are phase locked and synchronized across the whole chain. 
As an example, in Figs.~\ref{fig:Fig3}(b) and \ref{fig:Fig3}(e) we choose a inversion symmetric initial state.

The synchronization described so far only involve a single frequency in the system.
However, there also exists the case where the left and right ends of the chain oscillate with different frequencies.
Moreover, the oscillation amplitudes differ on both sides of the chain.
This is the case for $\phi\in [0.58\pi,2\pi/3)$ or $\phi \in (2\pi/3,0.75\pi]$ and an example is shown in Fig.~\ref{fig:Fig3}(c) for the case of of $\phi = 0.7\pi$.
Nevertheless, shuttles belonging to the same end of the chain are still synchronized and phase locked. 

As the self-oscillation of an electron shuttle relies on the interplay of electron tunneling and mechanical motion of the oscillator, the mechanical synchronization of the chain has a direct consequence on the dot occupation $q_l$ of each shuttle.
%
%The electron current $I_l(t)$ of each shuttle $l$ is defined as \cite{WaechtlerEtAlNJP2019}
%\begin{equation}
%I_l(t) = \Gamma_\text{out}[x_l(t)]\left[1- q_l(t)\right] -\Gamma_\text{in}[x_l(t)] q_l(t). 
%\end{equation}
%%
For visualization purposes, we again show in Figs.~\ref{fig:Fig3}(d)--\ref{fig:Fig3}(f) the deviation $\delta q_l(t) = q_l(t) -\langle q_l\rangle$. 
If the shuttles are oscillating, the dot occupation is altered in time due to the exponential dependency of the tunneling rates [see Eq.~(\ref{eq:Rates})]. 
For the examples we have discussed above [Figs.~\ref{fig:Fig3}(a)--\ref{fig:Fig3}(c)], we show the corresponding modulations of the dot occupation in Figs.~\ref{fig:Fig3}(d)--\ref{fig:Fig3}(f), which clearly show the same synchronization patterns as the mechanical motion.
As the dot occupation may be observed directly by nearby quantum point contacts, the synchronization may be probed directly with current technology \cite{FujisawaEtAlScience2006, FlindtEtAlPNAS2009}.
Note that $0 \leq q_l(t) \leq 1$ for all times and only the deviations $\delta q_l(t)$ may become negative.

We have discussed three different dynamical scenarios of synchronization present in the system depending on the global phase $\phi$.
As only the ends of the chains are excited the underlying topology of the chain influences synchronization. 
For a better understanding of this interplay of topology and synchronization, we investigate the system using a linear stability analysis in the next section.

\section{Linear stability analysis}\label{sec:LinearStability}

Linear stability analysis has been proven a useful concept to investigate the emergence of stable periodic motions in nonlinear systems~\cite{Strogatz2014}. 
To perform this analysis in the system at hand of coupled electron shuttles, we work in the collective mode basis $\bar{\mathbf X}$ such that the dynamics is described by $\dot{\bar{\mathbf X}} = \mathbf f(\bar{\mathbf X})$ [see Eq.~(\ref{eq:DynamicsNormalMode})]. 
We expand this nonlinear equation around the fixed point $\bar{\mathbf X}_\text{fix}$ with $\mathbf f(\bar{\mathbf X}_\text{fix}) \equiv 0$, which needs to be found numerically, up to first order, i.e., 
\begin{equation}
\dot{\bar{\mathbf X}} \approx \mathbf J\left(\bar{\mathbf X}_\text{fix}\right)\left(\bar{\mathbf X}-\bar{\mathbf X}_\text{fix}\right).
\end{equation}
Here, $\mathbf J(\bar{\mathbf X}_\text{fix})$ denotes the Jacobian matrix with entries $\mathbf J_{ij} = \partial \mathbf f_i/\partial \bar{\mathbf X}_j$ and eigenvalues $z_i$ evaluated at $\bar{\mathbf X}_\text{fix}$. 

\begin{figure}
\begin{center}
\includegraphics[width=\columnwidth]{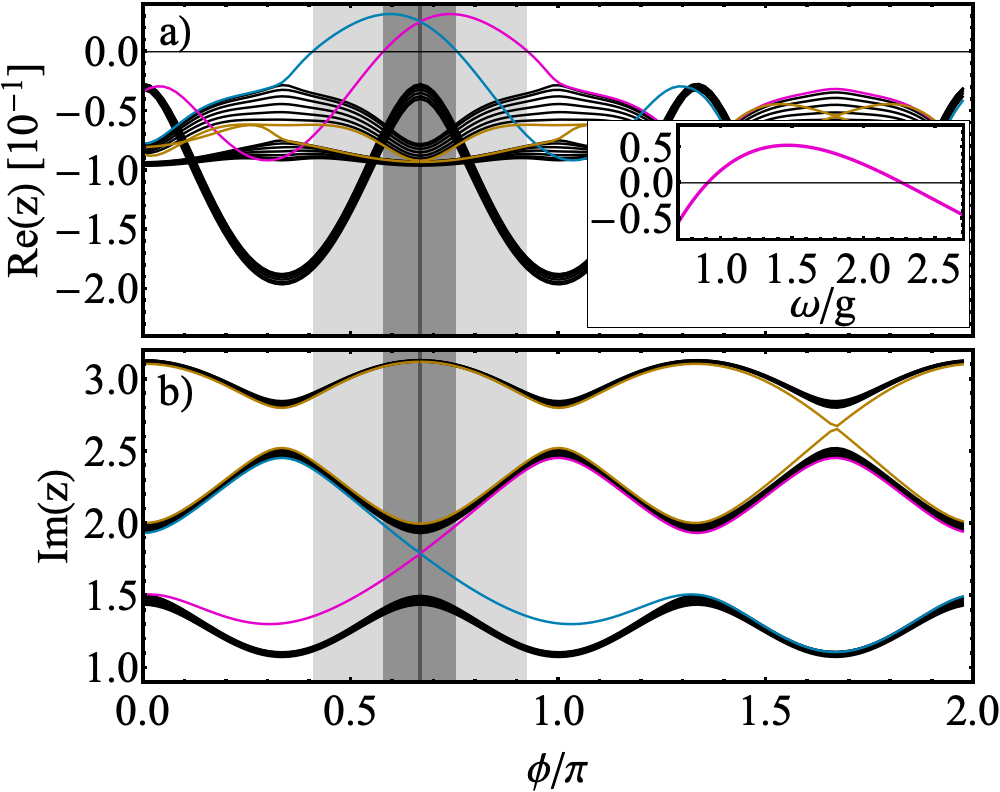}
\end{center}
\caption{Linear stability analysis: Real (a) and imaginary part (b) of eivenvalues $z_i$ of the Jacobian $\mathbf J(\bar{\mathbf X}_\text{fix})$ evaluated at the fixed point. As the global phase $\phi$ is varied, collective edge states are excited if $\text{Re}(z_i)>0$. The corresponding oscillation frequency is given by $\text{Im}(z_i)$. In the light gray shaded area, only one edge state is excited (blue right and pink left), whereas in the dark gray shaded area, both edge states are excited in general with different oscillation frequencies. At $\phi=2\pi/3$, the oscillation frequencies cross in panel (b), such that both edge states are synchronized. Inset: Real part of $z_i$ for a single shuttle as function of the oscillator frequency. Parameters: $\beta V = 150.0$, $\alpha \lambda = 0.06$, $\Delta / g = 1.0$, $\Gamma / \gamma = 1.0$.}
\label{fig:Fig4}
\end{figure}

The real part of the eigenvalues, $\text{Re}(z_i)$, gives us information about the stability of the fixed point $\bar{\mathbf X}_\text{fix}$. 
As long as $\text{Re}(z_i) < 0$ for all $i$, all solutions of the dynamical system $\dot{\bar{\mathbf X}}=\mathbf f(\bar{\mathbf X})$ are attracted into the fixed point $\bar{\mathbf X}_\text{fix}$ for long times. 
However, the fixed point becomes linearly unstable if $\text{Re}(z_i)> 0$ for at least one $i$. 
Then, the corresponding collective mode undergoes a Hopf bifurcation into a stable limit cycle with periodic motion and $\text{Im}(z_i)$ determines the oscillation frequency of this mode ~\cite{Strogatz2014}.

We show the relevant excerpts of the real part [Fig.~\ref{fig:Fig4}(a)] and imaginary part [Fig.~\ref{fig:Fig4}(b)] of the eigenvalues of $\mathbf J(\bar{\mathbf X}_\text{fix})$ as function of the global phase $\phi$, where each $z_i$ comes as a conjugated pair.
We indicate the left (pink) and right (blue) edge states accordingly to Fig.~\ref{fig:Fig2}(b). 
As $\phi$ is varied, we can understand the different synchronization scenarios discussed in Sec.~\ref{sec:Synchronization} by means of the linear stability analysis. 
First, for $\phi < 0.41\pi$ and $\phi > 0.92\pi$ (white area in Fig.~\ref{fig:Fig4}), the real part of all eigenvalues is negative, such that all collective modes are linearly stable and fall into the fixed point for long times. 
This corresponds to the trivial situation, where all shuttles are at rest. 

For $\phi \in (0.41\pi, 0.58\pi)$ or $\phi \in (0.75\pi,0.92\pi)$ (light gray area Fig.~\ref{fig:Fig4}), the real part of exactly one edge state is positive. 
Hence, this edge state, located either at the right (blue) or left (pink) end of the chain, becomes linearly unstable and all shuttles belonging to this edge state oscillate synchronously with a frequency according to the frequency in Fig.~\ref{fig:Fig4}(b). 
If $\text{Re}(z_i)>0$ for both edge states (dark gray area in Fig.~\ref{fig:Fig4}), both edges become unstable. 
In general, the two ends of the chain will oscillate with different frequencies as seen by the imaginary part of the eigenvalues, Fig.~\ref{fig:Fig4}(b).
Thus, two shuttles located at different sides of the chain oscillate with different frequencies and cannot be phase locked. 
However, for the special case of $\phi=2\pi/3$, the two branches cross and the frequencies of the edge states coincide.
Then, all shuttles oscillate synchronized and phase locked as discussed in Sec.~\ref{sec:Synchronization}.

For a better understanding of why the edges of the chain are excited while the bulk shuttles are at rest, we investigate the influence of the oscillator frequency on the self-oscillation of a single-electron shuttle described by Eq.~(\ref{eq:NonLinearSingleShuttle}).
In the inset of Fig.~\ref{fig:Fig4}(a), we show $\text{Re}(z_i)$ of the linear stability analysis for a single shuttle as function of the oscillator frequency $\omega$. 
For $\omega < 0.9 g$ and $\omega > 2.3 g$, the fixed point is linearly stable and the shuttle is at rest in the long time limit. 
However, for $0.9 g \leq \omega \leq 2.3 g$ the stability of the fixed point changes, leading to self-oscillation of the system. 
Moreover, the largest values of $\text{Re}(z_i)$ are centered around $\omega \approx 1.5 g$. 
We therefore argue in the case of a chain of electron shuttles, which in the collective mode basis is described by Eq.~(\ref{eq:DynamicsNormalMode}), that collective states with frequencies similar to the self-oscillation frequency of a single shuttle are excited. 
Examining Fig.~\ref{fig:Fig2}(b), we find that this is the case for the midgap edge states located between the first and second bands.

We can interpret this as follows: The applied bias voltage $V$ is just large enough to excite collective modes with matching frequencies located at the ends of the chain. 
If the inversion symmetry of the trimer chain is preserved for $\phi=2\pi/3$ (see Sec.~\ref{subsec:Topology}) and the edge states are topological, long-ranged correlations along the chain lead to synchronization at the ends of the chain. 
On the other hand, if the edge states are not of topological nature, they oscillate with different frequencies.
Hence, the underlying topology manifests itself in the dissipative nonequilibrium system as synchronized motion of the chain ends.
If $V$ is increased further, not only the edge states but also bulk states with additional frequencies would be excited, destroying the synchronization observed here.
Note that by changing system parameters, for example, the bare tunneling rate $\Gamma$, the excitable frequencies of the single shuttle can be shifted. We then expect that the midgap edge states located between the second and third bands can be excited instead of the ones located between the first and the second bands.

\section{Robustness against disorder}\label{sec:Disorder}

\begin{figure}
\begin{center}
\includegraphics[width=\columnwidth]{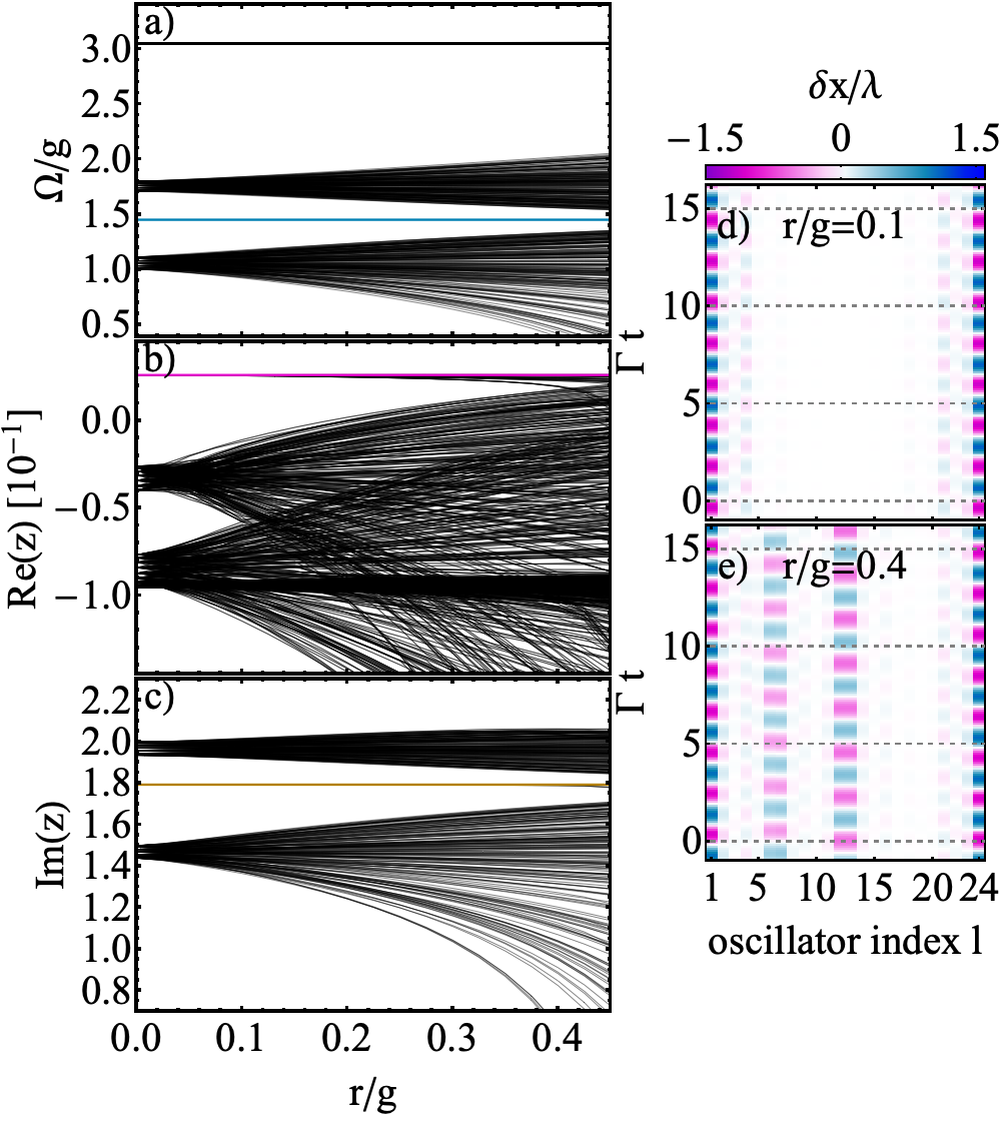}
\end{center}
\caption{(a) Frequency spectrum of the trimer chain of $N=24$ harmonic oscillators for $\phi = 2\pi/3$ [vertical line in Fig.~\ref{fig:Fig4}(a)] with increasing amount of disorder $r$ for $30$ realizations of disorder. The edge state (blue) is topologically protected against the disorder for large values of $r$. [(b)--(c)] $\text{Re}(z_i)$ and $\text{Im}(z_i)$ of the Jacobian of the linear stability analysis with increasing amount of disorder. The topologically protected edge state located at both ends of the chain persists for large values of $r$; however, additional bulk states with different frequencies may become linearly unstable. [(d)--(e)] Oscillating position $\delta x_l(t)$ of the shuttles for one realization of disorder with $r/g = 0.1$ and $r/g = 0.4$, respectively, and inversion symmetric initial states. Parameters: $\beta V = 150.0$, $\alpha \lambda = 0.06$, $\Delta / g = 1.0$, $\Gamma / \gamma = 1.0$.}
\label{fig:Fig5}
\end{figure}

From the discussion about the topology of trimer chains of harmonic oscillators in Sec.~\ref{subsec:Topology}, we know that not all edge states are of topological nature. 
They are topological only if for $\phi = 2\pi/3$ and $\phi = 5\pi/3$ the inversion symmetry is preserved and the edge states are protected against local disorder. 
We first discuss the topological protection in the trimer chains of harmonic oscillators alone.
The inversion symmetric case relevant for the discussion of synchronization of electron shuttles is described by the Hamiltonian $H_\text{osc}(2\pi/3)$ [see Eq.~(\ref{eq:HamiltonianTopology})].
Here, we consider spatial disorder, which preserves the intratrimer symmetry, and thus we only allow for disorder in the coupling between trimers. 
To be specific, the nanomechanical coupling between oscillators of units $A$ and $C$ is given by $(g+\delta g_i)^2$, where $\delta g_i$ is a uniformly distributed random number $\delta g_i \in \{-r,r\}$.
In Fig.~\ref{fig:Fig5}(a), we show the frequency spectrum of $H_\text{osc}(2\pi/3)$ for increasing amount of disorder $r$ for $30$ realizations. 
We can observe that even if details of the spectrum are modified through the disorder, the midgap state (blue) is topologically protected for very large amounts of disorder. 

Motivated by the previous observation, we are interested if synchronized edge states of the chain of electron shuttles are similarly protected by topology. 
To this end, we investigate the influence of the same disorder as above on the eigenvalues $z_i$ of the Jacobian of the linear stability analysis discussed in Sec.~\ref{sec:LinearStability}.
In Figs.~\ref{fig:Fig5}(b) and \ref{fig:Fig5}(c), we show $\text{Re}(z_i)$ and $\text{Im}(z_i)$ for $30$ realizations of disorder with increasing strength $r$. 
We first observe, that the frequency of the synchronized edge state [yellow in Fig.~\ref{fig:Fig5}(c)] is unaffected, similarly to the trimer chain of harmonic oscillators. 
However, in the spectrum of $\text{Re}(z_i)$, which accounts for the emergence of stable collective motion in the systems, is highly altered by the disorder.
First, for weak disorder strength $r$, only the synchronized edge states are linearly unstable ($\text{Re}>0$, pink), such that the synchronization is equivalent to no disorder. 
This can also be seen in Fig.~\ref{fig:Fig5}(d), where we show again $\delta x_l(t)$ for one realization of disorder such that $r/g = 0.1$. 

On the other hand, for large amounts of disorder, the real part of additional eigenvalues are positive.
The corresponding collective modes are bulk states.  
Hence, shuttles belonging to the middle of the chain oscillate.
The associated frequencies, however, are different from the edge states, such that shuttles at the ends and in the middle of the chain are not synchronized [see Fig.~\ref{fig:Fig5}(e)].
Conversely, the ends of the chain still perform synchronized motion. 
However, $\text{Re}(z_i)$ may be different for the two edges. 
Then, the edges oscillate with different amplitudes and are out of phase even with inversion symmetric initial states, as can be seen in Fig.~\ref{fig:Fig5}(e).
Hence, the synchronized motion of the shuttles located at the ends of the chain is topologically protected whereas the disorder manifests itself in breaking the symmetry of the oscillation amplitudes and phase and by the appearance of unsynchronized bulk states. 

\section{Summary and conclusions}\label{sec:Conclusions}

In summary, we discuss in this work a system which connects synchronization with topology in a dissipative nonequilibrium setup. 
Motivated by models of condensed matter physics we introduce topology into the system of coupled electron shuttles by modulation of the oscillation frequency of each shuttle. 
The obtained trimer chain exhibits synchronized motion located at the ends of the chain.
By investigation of a linear stability analysis of the nonlinear dynamical system, we explain the emergence of synchronized states by means of the underlying topology. 
If the inversion symmetry is preserved, longranged correlations lead to synchronized motion of both ends of the chain.
However, if the observed edge states are not of topological nature both ends of the chain oscillate with different frequencies or one end does not oscillate at all. 
Hence, the synchronization of the shuttles is a direct consequence of the topology of the trimer chain.
Moreover, the synchronization of the ends is topologically protected against local disorder, which preserves the intratrimer symmetry.
However, the local disorder manifests itself in breaking the symmetry of the oscillation amplitudes and phase at the ends of chain and by the appearance of unsynchronized bulk states.
The synchronized movement of the shuttles may be directly probed by the dot occupation of each shuttle, which shows the same synchronized periodicity as the shuttles itself.

As the symmetry protects the synchronization of the electron shuttles, we also expect the synchronized edge states to persist if thermal fluctuations of the nanomechanical oscillators are taken into account or by considering a fully quantum description of the shuttles.
Beyond this, it would be very interesting to explore the effects of well known phenomena from the field of topology on synchronization.
Here, one may extend the chain to a two-dimensional grid to investigate synchronized traveling edge states or topologically pump the synchronized edge states by varying the global phase $\phi$ over time.
Also, exploring the coexistence of synchronized and unsynchronized states, so-called chimera states, which are protected by topology, are of great interest.
The presented system thus may serve as a test bed not only theoretically but also experimentally for very interesting physics to be investigated and opens another avenue to explore topological protection in synchronized systems.

\begin{acknowledgements}
We are thankful for enlightening discussions with S. B\"ohling. V. M. B. acknowledges fruitful discussions with W. k. Mok and H. Heimonen. We gratefully acknowledge financial support from the Deutsche Forschungsgemeinschaft (DFG, German Research Foundation) through Project No. BR1528/8-2. This work was supported in part by the MEXT Quantum Leap Flagship Program (MEXT Q-LEAP) Grant No. JPMXS0118069605 and the JSPS KAKENHI Grant No. 19H00662.
\end{acknowledgements}

\end{document}